\documentclass[%
reprint,
 prl,            
 amsmath,amssymb,
 aps
]{revtex4-2}

\usepackage{graphicx}

\begin{document}

\title{Energy Bunching from Sub-Cycle Ionization Injection\\ in Laser Wakefield Acceleration}
\author{A.~Angella}
\email{andrea.angella@fysik.lu.se} 
\author{E.~Löfquist}
\author{C.~Gustafsson}
\author{V.~Poulain}
\author{F.~D'Souza}
\author{C.~Guo}
\author{A.~Persson}
\author{P.~Eng-Johnsson}
\author{C.-G.~Wahlström}
\author{O.~Lundh}
\email{olle.lundh@fysik.lu.se} 
\affiliation{Department of Physics, Lund University, P.O. Box 118, SE-22100 Lund, Sweden}
\date{August 14, 2025}

\begin{abstract}
We report the first experimental observation of carrier-envelope phase-driven energy bunching in laser wakefield acceleration. Using a few-cycle ($\sim$9~fs), multi-terawatt laser pulse and ionization injection in a helium–nitrogen gas mixture, we observe electron spectra composed of multiple quasi-monoenergetic peaks with regular narrow energy spacing. This comb structure arises from intermittent injection from successive half-cycles of the laser field, enabled by the evolving carrier-envelope phase during propagation in the plasma. These findings establish sub‑cycle ionization injection as a potential route to attosecond control in plasma acceleration, enabling injection and beam structuring synchronized to the optical waveform on sub-femtosecond timescales.
\end{abstract}

\maketitle

Laser wakefield acceleration (LWFA), first proposed by Tajima and Dawson \cite{Tajima1979}, enables compact, high-gradient acceleration of electrons to relativistic energies, with accelerating fields orders of magnitude stronger than those achievable with conventional accelerator technology. A central challenge is controlling the injection of electrons into the accelerating structure of the plasma. Among the controlled schemes, ionization injection, first demonstrated in helium–nitrogen mixtures \cite{Pak2010,McGuffey2010}, produces stable, high-charge beams by releasing inner-shell electrons only near the peak of the laser field. However, when this process extends over many optical cycles, it inevitably yields large energy spreads, motivating approaches that can confine injection to ultrashort, potentially sub-cycle, timescales.

Few-cycle, multi-terawatt lasers open such a regime, where the instantaneous electric field, not just the pulse envelope, governs plasma dynamics. In this regime, the carrier-envelope phase (CEP) is critical. Simulations by Lifschitz and Malka \cite{Lifschitz2012} predicted that CEP evolution during propagation shifts the timing of electron release at the half-cycle level, imprinting the laser waveform onto the electron beam phase space and producing a comb-like energy spectrum. Building on this idea, theory and simulations have predicted CEP‑dependent ultrashort electron bunches \cite{Zhang2022,Xu2016}, sub‑femtosecond injection and asymmetric plasma dynamics \cite{Nerush2009CEP,Kim2021,Huijts2021}, and experiments with mJ‑class, kHz lasers have revealed phase-sensitive electron beam pointing \cite{Huijts2022,Rovige2023,Seidel2024}. 

\begin{figure}[b]
    \includegraphics[width=\linewidth]{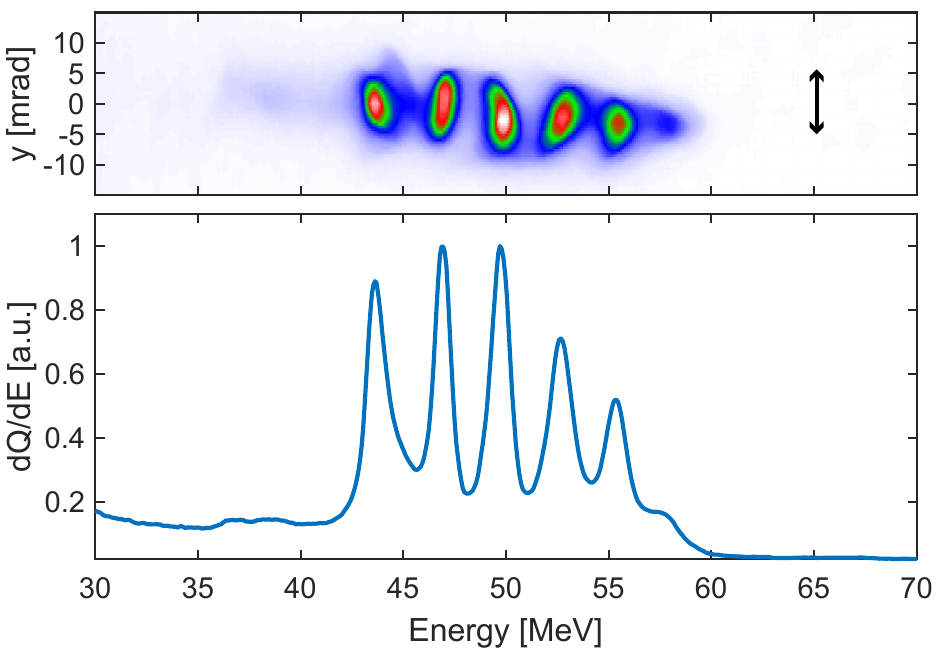}
    \caption{\label{fig:1} Typical single-shot electron energy spectrum from ionization injection in the few-cycle regime, exhibiting a comb of quasi-monoenergetic peaks with regular spacing. The double arrow indicates the direction of the laser polarization.}
\end{figure}

Here we present the first experimental observation of CEP-driven energy bunching in LWFA using a few-cycle laser pulse and ionization injection in a helium-nitrogen mixture. 
We observe a striking energy comb in the electron spectrum (Fig.~\ref{fig:1}) consisting of narrow, quasi-monoenergetic peaks with regular spacing, a signature of intermittent injection from successive half-cycles of the laser field. As the CEP evolves during propagation, field extrema alternately cross the ionization threshold for the K-shell of nitrogen, releasing electrons at distinct positions. Each bunch then experiences a different acceleration length, yielding the observed energy comb. This establishes that sub-cycle ionization injection is sensitive to the evolving CEP, providing a potential route toward attosecond electron bunch generation in plasma accelerators.

Experiments were performed with LUCID, the multi-terawatt OPCPA laser system at the Lund Laser Centre \cite{Balciunas2024}, delivering 9~fs pulses and 250~mJ before transport, corresponding to 190~mJ on target, at a central wavelength of 850~nm and 10 Hz repetition rate.  The system includes active stabilization of the carrier-envelope phase, using an $f$-2$f$ interferometer located upstream of the compressor. The spatial profile was corrected with a deformable mirror, and the beam was focused by an \emph{f}/15 off-axis parabolic mirror into a supersonic gas jet from a 1.5~mm pulsed nozzle, with focus 2~mm above the orifice. The target consisted of a 99:1 mixture of helium and nitrogen. The plasma density, $n_e$, was determined interferometrically using a wavefront sensor and assuming full ionization of helium and the five outermost electrons of nitrogen.

The energy bunching reported here was only observed in a narrow electron density window around (1.5-1.6)$\times 10^{19}\,\mathrm{cm}^{-3}$. The focal spot measured at full width at half maximum (FWHM) was 15~µm, with an inferred peak intensity of \(4\times 10^{18}\,\mathrm{W/cm^2}\) corresponding to a normalized vector potential of $a_0=1.5$.

Electrons accelerated in the wakefield were dispersed in the horizontal plane, perpendicular to the laser polarization, by a 0.83~T permanent dipole magnet onto a phosphor screen (Lanex regular). The fluorescence was imaged with a 16‑bit sCMOS camera. The energy calibration was obtained from particle tracking in the measured geometry of the nozzle, magnet, and detection plane, and the screen response was referenced to the charge calibration of Kurz \emph{et al.} \cite{Kurz2018}. The spectrometer provides an energy resolution of approximately 1\% at 50 MeV for a typical FWHM beam divergence of 5 mrad.

Optimization of the occurrence of energy bunching was performed by adjusting jet position, backing pressure, laser energy, and group delay dispersion (GDD), the latter controlled with an acousto‑optic programmable dispersive filter.  

The comb-like energy structure is consistently observed in our experiment. Figure~\ref{fig:1} shows a representative single-shot electron spectrum, with five narrow, regularly spaced peaks centered around 50 MeV. Each peak has a width of about 1 MeV (FWHM) and spacing of $\sim$3 MeV. To understand the origin of this structure, we analyze the evolution of the CEP during laser propagation in plasma, which determines the timing and location of successive injection events.

The origin of these energy combs lies in the continuous slip of the optical phase during laser propagation in plasma. Because the group velocity $v_g$ of the pulse envelope is below $c$ while the phase velocity $v_\phi$ of the optical carrier is above, the CEP evolves along the propagation axis. The slippage length over which the CEP advances by $2\pi$ is
\[
L_{2\pi} = \frac{\lambda_0c}{(v_\phi-v_g)}\simeq\lambda_0\frac{n_c}{n_e},
\]
where $\lambda_0$ is the laser central wavelength, $n_c$ is the critical plasma density, and the approximation assumes $n_c$~$\gg$~$n_e$. At $n_e = 1.6 \times 10^{19}\,\text{cm}^{-3}$ this gives $L_{2\pi} \simeq 80$~µm, so half-cycle-separated injection events are separated by $L_{2\pi}/2 \simeq 40$~µm. Each event releases electrons at a different position along the propagation axis and electrons released earlier experience a longer acceleration distance and therefore reach higher energies. The observed energy spacing of $\sim$3~MeV between peaks therefore reflects the average energy gain over $\sim$40~µm, corresponding to an effective accelerating field of $\sim$75~GV/m, consistent with expected values in this regime.

Crucially, this mechanism relies on the laser field remaining close to the ionization threshold for nitrogen K-shell electrons, such that only selected field extrema trigger injection. As a result, the number and regularity of energy peaks are highly sensitive to the evolution of laser intensity along the plasma, with small variations affecting how many injection events are gated and sustained.

\begin{figure}[t]
    \includegraphics[width=\linewidth]{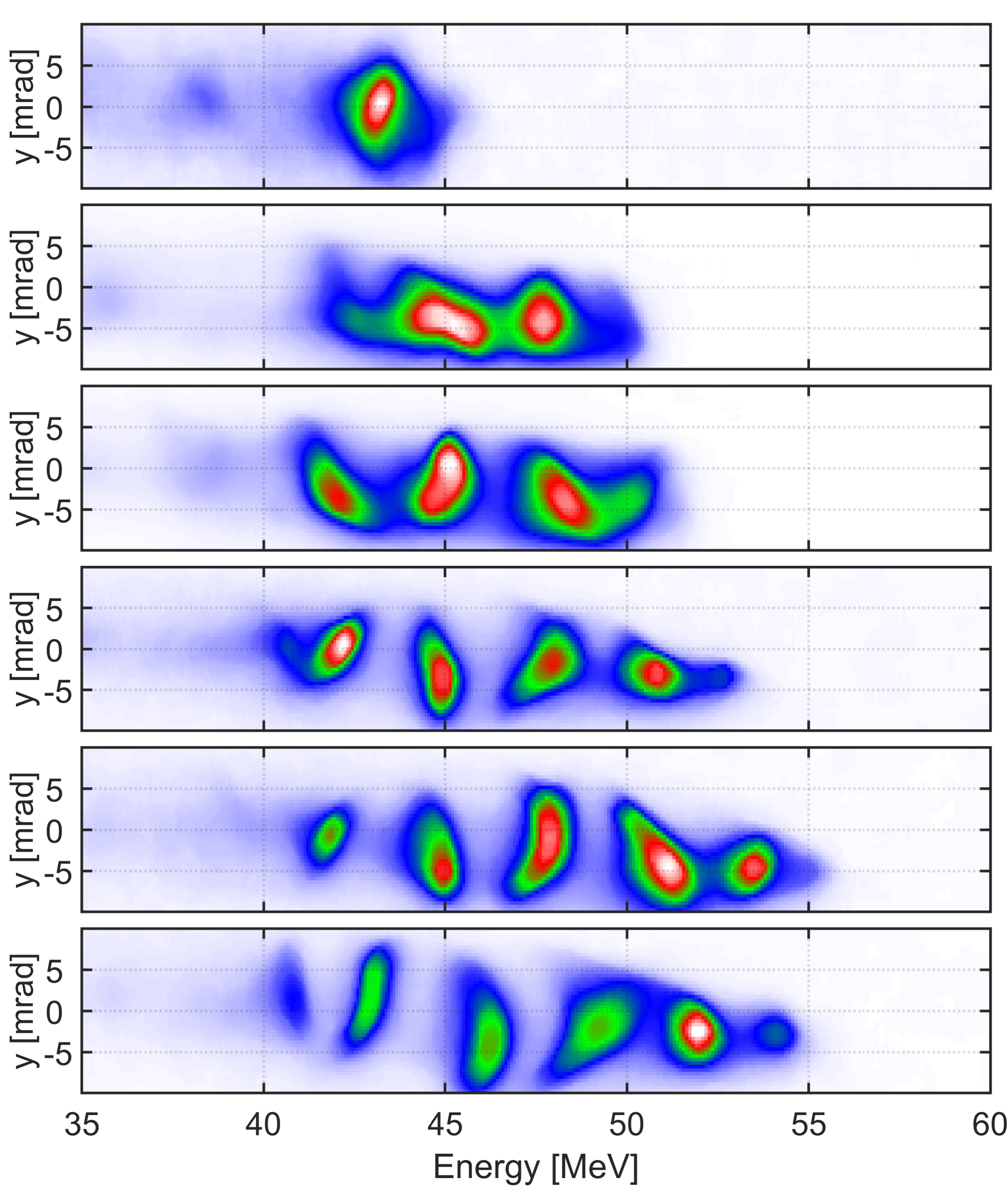}
    \caption{\label{fig:2} Examples of single-shot spectra, showing one to six energy-separated peaks, illustrating the variation observed in the measurements. Each spectrum is normalized to its maximum.}
\end{figure}

\begin{figure}[t]
    \includegraphics[width=\linewidth]{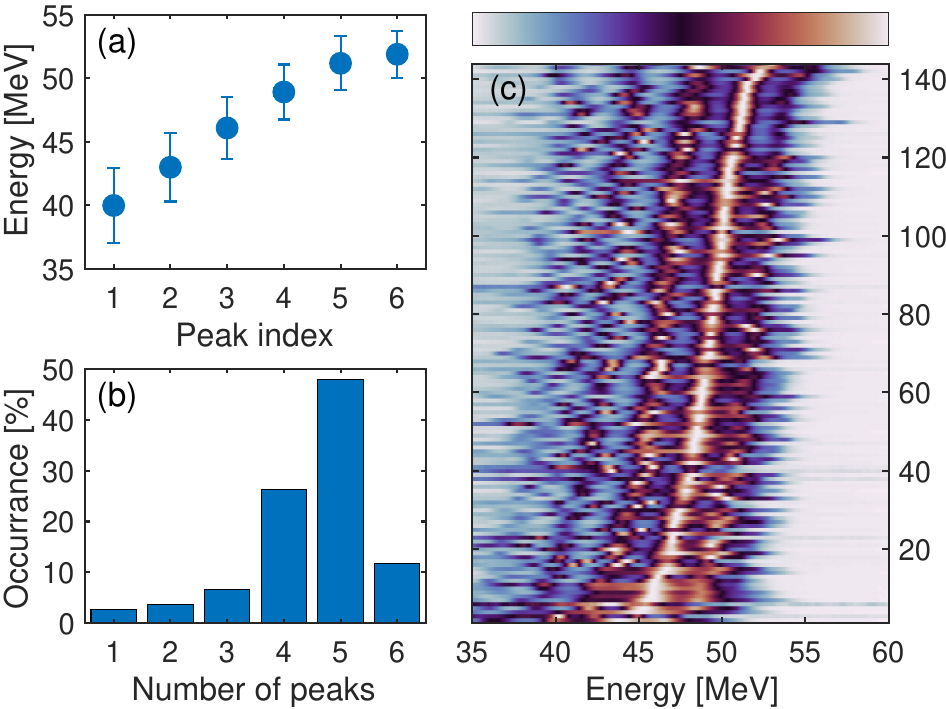}
    \caption{\label{fig:3}
    Statistical analysis of 300 single-shot spectra.
    (a)~Energies of resolved spectral peaks, indexed by increasing energy within each spectrum. Error bars show the standard deviation for each index.
    (b)~Distribution of the number of peaks observed per shot. 
    (c)~Stacked normalized energy spectra from all measurements with five resolved peaks, sorted by the fourth peak energy.}
\end{figure}

Figure~\ref{fig:2} shows example spectra with one to six quasi-monoenergetic peaks. More peaks appear when the intensity remains near threshold over a longer distance, enabling multiple injection events. This requires the field strength to be consistently close to threshold, high enough to ionize, but not so high to continuously inject. Small changes in plasma density, dispersion, or focusing conditions can shift this balance, making the process highly sensitive to initial conditions. As a result,  shot-to-shot variations arise even under nominally identical laser parameters. Statistical analysis of 300 shots (Fig.~\ref{fig:3}) reveals a monotonic increase in peak energy with peak index (Fig.~\ref{fig:3}a), and a strong preference for five-peak spectra (Fig.~\ref{fig:3}b). Peaks at higher energy correspond to electrons injected earlier and accelerated over longer distances. Bunching is observed in 98\% of shots, but disappears if the density is changed by more than 20\%, illustrating the sensitivity of the mechanism. No deterministic correlation was observed between the input CEP and the energy of the bunches, likely due to the sensitivity of the process to the pulse evolution. However, the relative energy spacing between bunches remains remarkably stable (Fig.~\ref{fig:3}c), consistent with intermittent injection at successive half-cycles of the laser field.

Importantly, no energy bunching is observed in pure helium, even under conditions where self-injection occurs, confirming that the effect depends on ionization injection. This distinguishes the observed energy combs from periodic structures predicted in simulations of self-injection in pure gases \cite{Xu2020}.

\begin{figure}[b]
    \includegraphics[width=\linewidth]{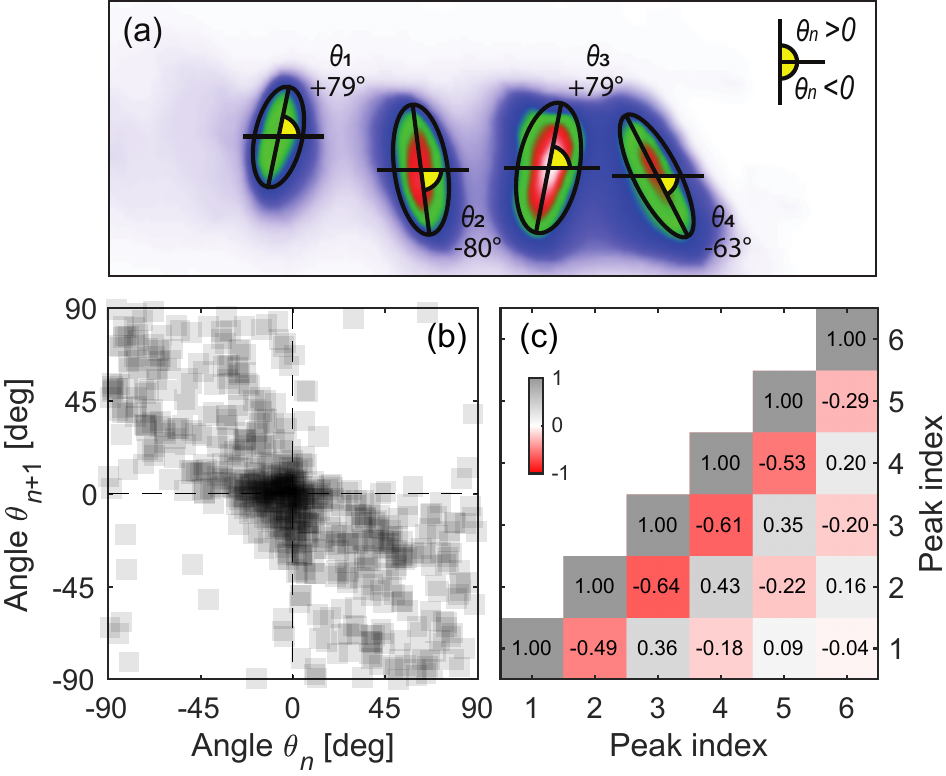}
    \caption{\label{fig:4} 
    (a)~Automated detection of individual electron bunches in Lanex screen images from 300 single shots, with ellipses fitted to each. The tilt of each ellipse's major axis quantifies the angular chirp. 
    (b)~Correlation between neighboring bunches reveals a clear pattern of alternating chirp direction, consistent with successive half-cycle injection.
    (c)~Correlation matrix between all angular pairs.}
\end{figure}

The energy peaks also show a characteristic angular chirp. Electrons born in alternating half-cycles receive opposite transverse kicks from the laser field, producing alternating tilts in the angle-energy space, directly visible on the scintillator screen. Tilt angles, extracted with an automated fitting algorithm, show that $\sim$80\% of adjacent bunches are anti-correlated in orientation (Fig.~\ref{fig:4}), consistent with successive half-cycle injection at optical extrema \cite{Lifschitz2012}. While the relative tilt reflects alternating transverse momentum at injection, the absolute deflection also depends on the dipole magnet strength, which maps longitudinal momentum to spatial displacement. Thus, the measured angular tilts encode both injection dynamics and spectrometer geometry. Although plasma evolution may stretch the bunch duration, the injection remains locked to individual optical half-cycles. This demonstrates that key aspects of plasma dynamics, including electron release, injection timing, and transverse momentum transfer, are governed by the instantaneous laser field on attosecond timescales.

\begin{figure}[t]
    \includegraphics[width=\linewidth]{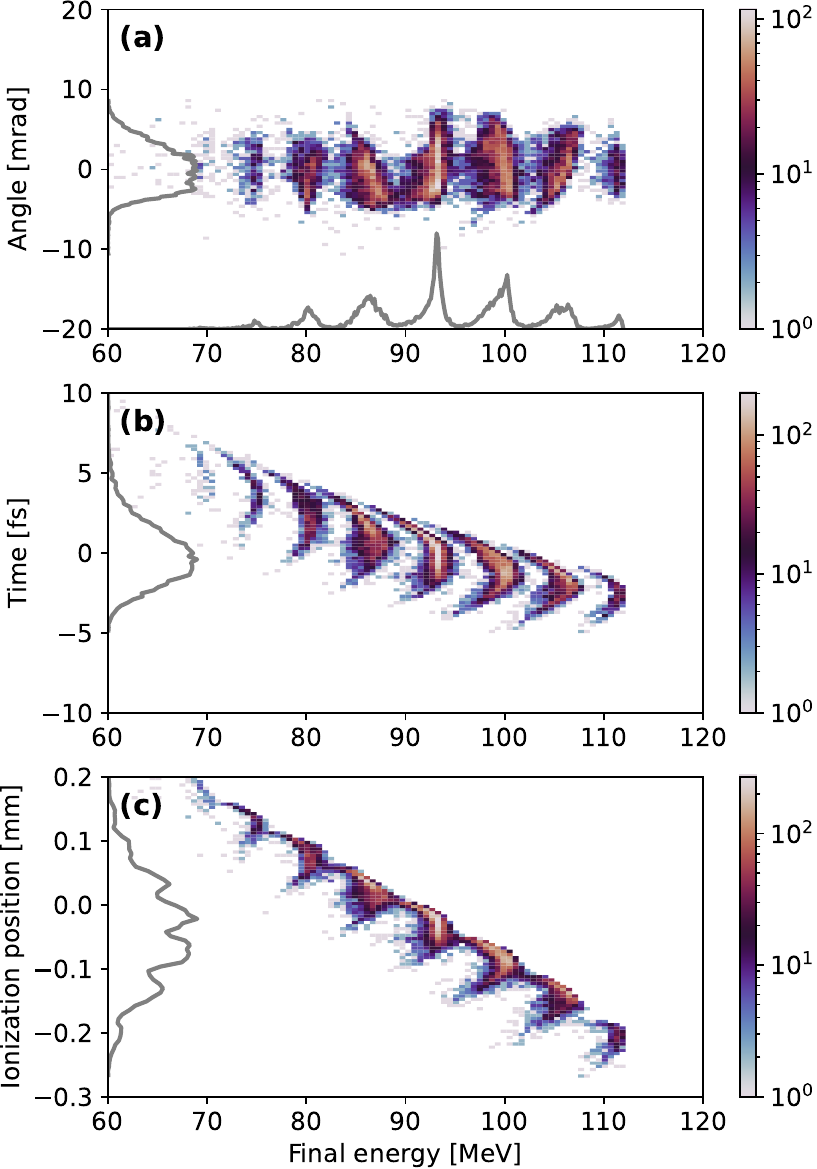}
    \caption{\label{fig:5} Particle distributions from a PIC simulation of sub-cycle ionization injection.
    (a)~Angle versus final energy, showing distinct energy bunching with alternating tilt angles. 
    (b)~Arrival time versus final energy, revealing overlapping longitudinal bunching. 
    (c)~Ionization position versus final energy, showing discrete injection events along the propagation axis.}
\end{figure}

To further support this interpretation, we performed 3D particle-in-cell (PIC) simulations using the open-source spectral code FBPIC~\cite{Lehe2016}, with parameters closely matching the experiment. The laser was modeled as a linearly polarized, 850~nm, 9~fs Gaussian pulse with spot size 15~µm and normalized vector potential $a_0=1.3$, propagating through a 2~mm-long plasma with a peaked $\cos^2$ density profile and electron density $1.2\times10^{19}\,\mathrm{cm^{-3}}$. The target consisted of a helium–nitrogen mixture, with ionization modeled using ADK rates~\cite{ADK1986}. The simulations reproduce energy bunching with an energy-dependent beam angle along the laser polarization axis (Fig.~\ref{fig:5}a), and a comb-like structure in longitudinal phase space (Fig.~\ref{fig:5}b). Injection occurs exclusively from nitrogen K-shell electrons. No self-injection is observed, and all histograms are computed from the nitrogen-born population. Discrete injection events appear along the plasma (Fig.~\ref{fig:5}c), with spacing consistent with the CEP slippage length. Simulated energies exceed those in the experiment, likely due to the use of an idealized Gaussian laser pulse, but the relative energy spacing ($\sim$6\%) is in good agreement with the measured spectra. Clear comb structures emerge only near threshold conditions, consistent with the narrow experimental window in plasma density. 

The close agreement of the observed energy bunching and alternating angular chirps with both theoretical predictions \cite{Lifschitz2012,Zhang2022} and supporting PIC simulations (Fig.~\ref{fig:5}) confirms that the half‑cycle structure of the laser field shapes the electron spectrum. This marks a fundamental departure from conventional ionization injection, which typically yields broad and continuous energy distributions. Sub-cycle control of the phase space opens a potential route toward attosecond-scale injection and the generation of isolated ultrashort electron bunches.

The restriction of energy bunching to a narrow density window is likely due to the combined effects of the plasma profile, the balance between slippage and injection lengths, and the intensity sensitivity of the mechanism. Injection requires the laser field to remain near the nitrogen K‑shell ionization threshold, and pulse evolution in plasma can easily shift the effective intensity away from this range.

Although demonstrated here at tens of MeV and using a 10~Hz laser, the mechanism is scalable to GeV‑class beams since the essential physics does not depend on the final energy or acceleration length. The same principles also apply at high repetition rates, where few‑cycle drivers operating in the kHz regime could support stable phase‑sensitive injection. These prospects open new possibilities for phase-driven electron beam tailoring at higher energies and higher repetition rates.

In conclusion, we report the first experimental evidence of CEP‑driven energy bunching in LWFA, where sub‑cycle injection synchronized to the laser waveform shapes  the electron spectrum. Looking ahead, tailoring the plasma profile could confine injection to a single half‑cycle, enabling energy-tunable, isolated sub‑femtosecond electron bunches, and offering attosecond control over injection timing in plasma accelerators.

\begin{acknowledgments}
This work was supported by the Swedish Research Council (VR 2019-04784, 2024-05698), the Knut and Alice Wallenberg Foundation (KAW 2019.0318, 2020.0111), and the EuPRAXIA-DN MSCA Doctoral Network (Horizon Europe, GA 101073480).
\end{acknowledgments}

\bibliography{bibliography}

\end{document}